\documentclass[12pt]{article}

\usepackage{epsfig,graphicx}
\usepackage{amsmath}
\usepackage{amssymb}
\usepackage{array}
\usepackage{delarray}
   \newcommand{\query}[1]{\marginpar{%
   \vskip-\baselineskip
   \raggedright\footnotesize
   \itshape\hrule\smallskip#1\par\smallskip\hrule}}
   \newcommand{\removequeries}{\renewcommand{\query}[1]{}}
   
\newcommand{\nc}{\newcommand}
\nc{\beq}{\begin{equation}}
\nc{\eeq}{\end{equation}}
\nc{\beqa}{\begin{eqnarray}}
\nc{\eeqa}{\end{eqnarray}}
\nc{\bea}{\begin{eqnarray}}
\nc{\eea}{\end{eqnarray}}
\nc{\ba}{\begin{array}}
\nc{\ea}{\end{array}}

\newcommand{\no}{\nonumber}

\newcommand{\OMIT}[1]{{}}

\newcommand\spur{\raise.15ex\hbox{/}\kern-.57em }

\newcommand{\cH}{\mathcal{H}}
\newcommand{\cO}{\mathcal{O}}
\newcommand{\cA}{\mathcal{A}}
\newcommand{\cL}{\mathcal{L}}

\newcommand{\vckm}{{V_{\text{CKM}}}}
\newcommand{\vdackm}{{V^\dagger_{\text{CKM}}}}
\newcommand{\upmns}{{U_{\text{PMNS}}}}
\newcommand{\udagpmns}{{U^\dagger_{\text{PMNS}}}}

\newcommand{\hc}{\text{h.c.}}

\newcommand{\lsim}{\mathrel{\hbox{\rlap{\hbox{\lower4pt\hbox{$\sim$}}}\hbox{$<$}}}}
\newcommand{\gsim}{\mathrel{\hbox{\rlap{\hbox{\lower4pt\hbox{$\sim$}}}\hbox{$>$}}}}
\newcommand{\yuu}{\lambda_u}
\newcommand{\ydd}{\lambda_d}
\newcommand{\yee}{\lambda_e}
\newcommand{\ynn}{\ynu}
\newcommand{\yuuph}{\lambda_u^{\phantom{\dagger}}}
\newcommand{\yddph}{\lambda_d^{\phantom{\dagger}}}
\newcommand{\yeeph}{\lambda_e^{\phantom{\dagger}}}
\newcommand{\ynnph}{\ynuph}
\newcommand{\ybu}{\bar\lambda_u}
\newcommand{\ybd}{\bar\lambda_d}
\newcommand{\ybe}{\bar\lambda_e}

\newcommand{\yfive}{\lambda_{5} }
\newcommand{\yten}{\lambda_{10} }
\newcommand{\ysig}{\lambda_{5}^\prime}
\newcommand{\ysf}{\lambda_{5}^{(\prime)}}
\newcommand{\ynu}{\lambda_{1}}
\newcommand{\yfiveph}{\lambda_{5}^{\phantom{\dagger}} }
\newcommand{\ytenph}{\lambda_{10}^{\phantom{\dagger}} }
\newcommand{\ysigph}{\lambda_{5}^{\prime\phantom{\dagger}} }
\newcommand{\ysfph}{\lambda_{5}^{(\prime)\phantom{\dagger}}}
\newcommand{\ynuph}{\lambda_{1}^{\phantom{\dagger}}}

\newcommand{\LYGUTa}{\cL^{(0)}_{\rm Y-GUT}}
\newcommand{\LYGUTb}{\cL^{({\rm 2H})}_{\rm Y-GUT}}
\newcommand{\LYGUTn}{\cL^{(\nu)}_{\rm Y-GUT}}
\newcommand{\XLFV}{\Delta^{(l)}}
\newcommand{\XQFV}{\Delta^{(q)}}
\newcommand{\xlfv}{\delta^{(l)}}

\def\npb#1#2#3{    {Nucl. Phys.}~B {\bf #1}, #3 (#2)}
\def\plb#1#2#3{    {Phys. Lett.}~B {\bf #1}, #3 (#2)}

\begin{document}

\removequeries


\title{{\bf Grand Unification and the Principle of \\  Minimal Flavor Violation}}
\vskip0.25in
\author{Benjam\'\i{}n Grinstein~$^a$, Vincenzo Cirigliano~$^{b,c}$, \\ 
Gino Isidori~$^d$, and Mark B. Wise~$^c$ \\[10 pt]
{\small $^a$~Department of Physics, University of California at San Diego, La Jolla, 
CA 92093, USA} \\
{\small $^b$~Theoretical Division, Los Alamos National Laboratory, Los Alamos NM 87545, USA} \\
{\small $^c$~California Institute of Technology, 452-48, Pasadena, CA 91125, USA} \\
{\small $^d$~INFN, Laboratori Nazionali di Frascati, Via E.~Fermi 40, I-00044 Frascati, Italy}}
\date{August 10, 2006}
\maketitle

\begin{abstract}
Minimal Flavor Violation is an attractive approach to suppress 
unacceptably large flavor changing neutral currents from  
beyond the standard model physics at the TeV scale. 
It can be used in theories with low energy supersymmetry, 
multi Higgs doublet theories and other extensions of 
the minimal standard model. 
We show how minimal flavor violation can be implemented 
in Grand Unified theories. 
\end{abstract}

\begin{flushright}
\vskip-6.1in
CALT-68-2606\\
UCSD/PTH-06-10\\
\end{flushright}

\setcounter{footnote}{0} \setcounter{page}{1}
\setcounter{section}{0} \setcounter{subsection}{0}
\setcounter{subsubsection}{0}

\newpage 
\section{Introduction} 

The minimal Standard Model (SM) contains three generations of quarks
in left-handed $SU(2)_L$ doublets, $Q_{iL}$, and right-handed $SU(2)_L$
singlets, $u_{iR}$ and $d_{iR}$\,, and three generations of leptons in
left-handed $SU(2)_L$ doublets, $L_{iL}$ and right-handed $SU(2)_L$
singlets $e_{iR}$.   The
only terms that break the $SU(3)_Q \times SU(3)_U\times SU(3)_D\times SU(3)_L\times
SU(3)_E$ flavor symmetry are the quark and lepton Yukawa couplings to
the Higgs doublet. Extensions of the standard model with new particles
at the TeV scale that couple to the quarks are severely constrained by
experimental limits on flavor changing neutral currents.  An
attractive way to prevent the new physics from generating unacceptably
large flavor changing neutral currents is the principle of Minimal
Flavor Violation (MFV) which attributes all the breaking of the
$SU(3)_Q \times SU(3)_U\times SU(3)_D$ part of the flavor group to the quark
Yukawas~\cite{Georgi,HR,MFV}.  This principle has been implemented in
specific models, such as strongly-interacting theories \cite{Georgi},
low-energy supersymmetry \cite{HR,MFV}, and multi-Higgs models
\cite{MFV,MW}, but it can also be formulated in terms of a generic
non-renormalizable effective field theory valid around and below the
TeV scale \cite{MFV}. 

Neutrino masses can be incorporated into the SM by adding three
generations of right-handed singlet neutrinos, $\nu_{iR}$, to the
model. The flavor symmetry of the model is now extended by an
additional $SU(3)_\nu$ factor. If the Majorana mass of these right
handed neutrinos is large, when integrated out they generate dimension
five operators that give Majorana masses to the left handed,
non-singlet neutrinos. If, on the other hand, the right-handed
neutrinos are light, then they combine with the left handed neutrinos
into quasi-Dirac particles.  Since neutrino masses break the leptonic flavor
symmetry of the SM, one can consider extending the notion of MFV to
the neutrino sector. The extension is trivial in the case of quasi-Dirac
neutrinos, but not so for the case of heavy right handed
neutrinos. The latter has the added feature that the small neutrino
masses are generated by the see-saw mechanism.  The (non-trivial)
extension of the principle of minimal flavor violation to the lepton
sector has recently been discussed in~\cite{MLFV} (see also
\cite{Sasha}).

In Grand Unified Theories (GUT) \cite{GUT} the quarks and leptons
appear in the same representations and so implementing the principle
of minimal flavor violation requires some modifications.  Consider the
case where the three generations of fermions fall into $\bar {\bf 5}$,
${\bf 10}$ and ${\bf 1}$ representations of $SU(5)$. The $\bar {\bf
5}$ representations $\psi_i$ contain the $d^c_{iR}$ quark fields and
the lepton doublet fields $L_{iL}$, the ${\bf 10}$ representations
$\chi_i$ contain the quark doublet fields $Q_{iL}$, $u^c_{iR}$ and
the lepton fields $e_{iR}^c$. Finally the singlet ${\bf 1}$
representations $N_i$ contain the right-handed neutrinos
$\nu_{iR}$. Evidently in such a unified theory the $SU(3)_Q \times SU(3)_U\times
SU(3)_D\times SU(3)_L\times SU(3)_E\times SU(3)_\nu $ symmetry is reduced to
$SU(3)_{\bar 5} \times SU(3)_{10}\times SU(3)_1$ via the identifications
$SU(3)_Q\sim SU(3)_U^*\sim SU(3)_E^*\sim SU(3)_{10}$ and $SU(3)_D^* \sim SU(3)_L \sim
SU(3)_{ \bar 5}$ (and $SU(3)_\nu\sim SU(3)_1$).  In this
paper we show how to implement the principle of minimal flavor
violation in grand unified theories based on the above particle
content of the unifying gauge group $SU(5)$.

\section{Flavor group and irreducible symmetry-breaking terms}
\label{sect:two}

The starting point to define the irreducible sources of flavor-symmetry 
breaking in a $SU(5)$ GUT framework are the following renormalizable 
Yukawa interactions 
\beq
\LYGUTa  =   \yfive^{ij} ~ \psi^T_i \chi_j  H^*_{5}  + 
                           \yten^{ij} ~ \chi^T_i  \chi_j  H_{5} + \hc~,
\eeq 
where $H_5$ is a Higgs  field in the {\bf 5} of $SU(5)$ and
 we have explicitly indicated the $3 \times 3$ flavor 
indexes (and omitted the $SU(5)$ indexes). Imposing the invariance of 
$\LYGUTa$  under the  transformations  
\beq
\psi \to  V_{\bar 5}~  \psi~, \qquad 
\chi \to  V_{10}~ \chi~,  
\eeq
which define the $SU(3)_{\bar 5} \times SU(3)_{10}$ flavor-symmetry group, 
allows us to identify the following spurion transformation properties:
\beq
\yfive \to  V_{\bar 5}^* ~\yfive~ V_{10}^\dagger~, \qquad \yten \to  V_{10}^* ~\yten~ V_{10}^\dagger~.
\eeq
Projecting $\LYGUTa$ into the basis of 
$SU(3)_c \times SU(2)_L\times U(1)_Y$ light fields, or integrating 
out the heavy components of  $H_{5}$, leads to the standard model couplings
\beq
\LYGUTa \supset  \cL_{\rm Y-SM} =  
  \yuu^{ij} ~ {\bar u}^i_R  Q^j_L  h^*
+ \ydd^{ij} ~ {\bar d}^i_R  Q^j_L  h +\yee^{ij} ~ {\bar e}^i_R   L^j_L  h +\hc~,
\label{eq:LY_SM}
\eeq
where $h$ is the SM Higgs field and the Yukawa couplings 
satisfy the minimal GUT relations~\cite{GUT}
\beq
 \yuu \propto \yten~, \qquad \yddph \propto \yee^T \propto \yfiveph~.
\label{eq:yd=ye}
\eeq

A natural extension of this minimal set up is provided by the
introduction of two Higgs fields, a {\bf 5} and a $ \bf{\bar5}$, in
the $SU(5)$ Yukawa interaction, with appropriate assignments of $U(1)$
charges in order to avoid tree-level FCNCs:
\beq
\LYGUTb  =   \yfive^{ij} ~ \psi^T_i \chi_j  H_{\bar 5}  + 
             \yten^{ij} ~ \chi^T_i  \chi_j  H_{5} + \hc~,
\eeq 
In this case the low-energy Yukawa interaction becomes 
a two-Higgs doublet model of type-II. 
This choice allows us to change the relative normalization 
of the spurions $\yfive$ and  $\yten$, but does not modify
the proportionality between $\ydd$ and  $\yee^T$ 
in Eq.~(\ref{eq:yd=ye}).

The proportionality between $\ydd$ and  $\yee^T$ implies
the following relations (evaluated at the GUT scale):
\beq 
m_\tau = m_b~, \qquad   
\frac{m_\mu}{m_\tau}  = \frac{m_s}{m_b}~, \qquad 
\frac{m_e}{m_\mu} = \frac{m_d}{m_s}~.
\label{eq:m_ratios}
\eeq
The last of these relations is badly broken by the experimental 
values of fermion masses~\cite{GJ} 
and, as we will discuss in the next section, 
is very stable with respect to radiative corrections.  
As a result, we are forced to introduce 
new $SU(3)_{\bar 5} \times SU(3)_{10}$ spurions 
which break the relation $\yee^T \propto \ydd$.
A simple way to achieve this goal is provided by 
the introduction of the dimension-five operator \cite{PlackSlop}
(see also \cite{Enrico})
\beq 
\frac{1}{M} (\ysig)^{ij} ~ 
\psi^T_i \Sigma \chi_j H_{\bar 5} + \hc~,
\label{eq:LYGUTs}
\eeq 
where $M$ is a heavy scale ($M \gg M_{\rm GUT}$),
$\Sigma$ is a Higgs field in the adjoint representation of $SU(5)$,
and $\ysig$ is a new spurion 
(transforming as $\yfive$ under the flavor group).
The high-scale vev of $\Sigma$ breaks $SU(5)$ preserving $SU(2)_L\times U(1)_Y$,
\beq
\langle \Sigma \rangle = M_{\rm GUT}~ {\rm diag}(1,1,1,-3/2,-3/2)~,
\eeq
such that the GUT relations (\ref{eq:yd=ye}) are modified, thus:
\beq
 \yuu \propto \yten~, \qquad \ydd \propto \left(\yfive +  \ysig\right)~, \qquad  
 \yee^T \propto  \left(\yfive - \frac{3}{2} \ysig\right)~,
\label{eq:y_gut}
\eeq 
where we have redefined the spurion $\ysig$ incorporating the 
$M_{\rm GUT}/M$ suppression factor. Note that the Higgs
combination $\Sigma H_{\bar 5}$ appearing in Eq.~(\ref{eq:LYGUTs})
contains a {\bf 45} representation of $SU(5)$. Thus a completely equivalent 
result can be obtained with a single {\bf 45} Higgs field
--with appropriate electroweak-scale vev breaking 
$SU(2)_L\times U(1)_Y$-- and a renormalizable Yukawa interaction.
The higher-dimensional operator in Eq.~(\ref{eq:LYGUTs}) has the 
advantage of providing a natural explanation for the smallness
of $\ysig$ which is supported by experimental data.

Whether we start from $\LYGUTa$ or $\LYGUTb$, 
and whether we add the non-renorma\-li\-za\-ble term in Eq.~(\ref{eq:LYGUTs})
or a {\bf 45} Higgs Yukawa term, a successful description 
of quark and charged-lepton masses requires the introduction of three  
independent symmetry-breaking spurions. From this point of 
view, the situation is very similar to the non-GUT MFV case 
analyzed in \cite{MFV}.
However, the flavor structure of the theory is quite different 
with respect to the non-GUT case: the symmetry group is substantially 
smaller and this allows the symmetry-breaking sources 
to appear in more ways in the low-energy effective operators. 

There are also non-renormalizable operators that contribute to the
up-type quark mass matrix. For example,
\beq
\label{tenprime}
\frac{1}{M}(\lambda_{10}^{\prime})^{ij}\chi_i^T \Sigma \chi_j H_5.
\eeq
Dimensional analysis suggests such operators could contribute
significantly to the up-quark mass. Since there are no mass relations
for the up-type quarks we neglect such contributions in this
paper. This convenient assumption has the advantage of leaving the 
up-type quark mass matrix symmetric, which simplifies our analysis of
structures that give rise to flavor violation in the low energy
effective theory. As above, we could replace the $\Sigma H_5/M$ combination
by a {\bf45} Higgs which leads to a potentially large shift in
the up-type Yukawa
by an anti-symmetric matrix. We
again neglect this contribution for simplicity, leaving the 
up-type quark mass matrix symmetric.

\medskip 
As far as the neutrino sector is concerned, the most natural 
choice in this framework is the introduction of three 
$SU(5)$-singlet fields corresponding to the right-handed neutrinos.
In general, this implies the enlargement of the 
flavor group to $SU(3)_{\bar 5} \times SU(3)_{10} \times SU(3)_{1}$
and the inclusion of two new spurions:
\beq
\LYGUTn  =   \ynu^{ij} ~ N^T_{i} \psi_j  H_{5}  + 
              M_R^{ij} N^T_{i} N_{j} + \hc
\eeq 
Note that the spurion $M_R$ breaks both the flavor-symmetry subgroup
$SU(3)_{1}$ and the $U(1)_{\rm LN}$ associated to total lepton number.
In addition, the right-handed neutrino mass term breaks a discrete
symmetry under which all spinor fields $f$ transform according to $f\to
if$ and the Higgs field(s) transform as $H\to-H$. Hence it is
technically natural to take $M_R$ to be significantly smaller than
$M_{\text{GUT}}$. A second consequence of this discrete symmetry is
that $M_R$ cannot be radiatively generated, even though the spurion
combination $\lambda_1\lambda_5^*\lambda_{10}(\lambda_1\lambda_5^*)^T$ transform precisely like the
spurion $M_R$ under the flavor symmetries.  The third and most
interesting consequence of this discrete symmetry is that it
constrains the scale of the dimension five operator that gives rise to
the left-handed neutrino masses in the low energy effective field
theory. To see this, focus
on the case where the eigenvalues of $M_R$ are all much larger than
the scale of new physics $\Lambda$. In the low energy effective theory the
left-handed neutrino mass arises from a dimension-five operator
$hL_L^{T} L_L h$, where $h$ stands for the Higgs doublet and flavor
and gauge indexes have been suppressed.  The spurion analysis alone
allows as coefficient to this operator the combination
$\lambda_5\lambda^\dagger_{10}\lambda_5^T/ \Lambda$, but this is forbidden by the discrete symmetry
$L_L\to i L_L$, $h\to -h$.  On the other hand, since the spurion $M_R$
transforms as $M_R\to -M_R$, the coefficient $\lambda_1^TM_R^{-1}\lambda_1$ is
allowed by both flavor and discrete symmetries.  Hence the
dimension-five operator is necessarily suppressed by the large mass
scale $M_R$.

If neutrinos are quasi-Dirac particles then $\lambda_1$ and $M_R$ are very small,
$\lambda_1\sim M_R/v\lesssim 10^{-11}$, where $v$ is the SM Higgs doublet expectation
value. Hence, in the case that neutrinos are quasi-Dirac particles the
effect of these matrices on flavor physics is unobservably small, and
all new effects are codified in $\lambda_{10}$ and $\lambda_5^{(\prime)}$. This is not
the case if neutrinos are Majorana particles, in which instance the
right-handed neutrinos may be heavy. In fact, very heavy right-handed
neutrinos, with mass below $M_{\text{GUT}}$ but well above the TeV scale 
give rise automatically to light Majorana neutrinos, of mass $\sim |\lambda_1v|^2/M_R$,
through the see-saw mechanism. In this situation $\lambda_1$ may be large
enough to have sizable effects on flavor physics. This is explored
below in the following two sections where we largely focus on the
heavy right-handed neutrino case.

\medskip 
Summarizing, the transformation properties of irreducible spurions 
and low-energy fields are:
\beq
\ba{ccc}
Q_L &\to& V_{10}   ~Q_L \\
u_R &\to& V_{10}^* ~u_R \\
d_R &\to& V_{\bar 5}^* ~d_R \\
L_L &\to& V_{\bar 5}   ~L_L \\
e_R &\to& V_{10}^* ~e_R 
\ea
\qquad\qquad
\ba{ccc}
\yten  &\to&  V_{10}^{*\phantom{\dagger}}     ~\ytenph~ V_{10}^\dagger  \\
\yfive &\to&  V_{\bar 5}^{*\phantom{\dagger}} ~\yfiveph~ V_{10}^\dagger \\ 
\ysig  &\to&  V_{\bar 5}^{*\phantom{\dagger}} ~\ysigph~ V_{10}^\dagger \\
\ynu   &\to&  V_{1}^{*\phantom{\dagger}}   ~\ynuph~  V_{\bar 5}^\dagger  \\
 M_R   &\to&   ~V_{1}^{*\phantom{\dagger}}  ~M_R^{\phantom{\dagger}}~ V_{1}^\dagger  
\ea
\eeq

\newpage
\section{Quark and Lepton Masses}
\label{sect:QLMasses}

Taking into account radiative corrections, 
the effective low-energy (electroweak-scale) Yukawa 
couplings responsible for quark and charged-lepton masses 
assume the following form:
\begin{align}
  \yuu &= a_u \left[ \ytenph  + \epsilon_{1u}  \ytenph\yten^\dagger\ytenph 
                      + \epsilon_{2u}^{(\prime\prime)}  
  \ytenph(\ysfph)^\dagger\ysfph  + \ldots  \right]~, \no \\
  \ydd &= a_d \left[ \left(\yfiveph +  \ysigph\right) +  \epsilon^\prime_{d1}   \ysigph
                      + \epsilon^{(\prime)}_{d2} \ysfph\yten^\dagger\ytenph 
                      + \epsilon^{(\prime)}_{d3} \ynu^T \ynu^* \ysfph +  \ldots  \right]~,  \no \\
\yee^T &= a_e \left[ \left(\yfiveph - \frac{3}{2}  \ysigph\right)  +  \epsilon^\prime_{e1}  \ysigph
                      + \epsilon^{(\prime)}_{e2}  \ysfph\yten^\dagger\ytenph 
                      + \epsilon^{(\prime)}_{e3}  \ynu^T \ynu^* \ysfph  + \ldots  \right]~, 
\label{eq:y_eff}
\end{align}
where $\ysf$ denotes either $\yfive$ or $\ysig$. 
The leading $a_i$ coefficients encode potentially large 
QCD logarithms (in the case of the quark Yukawa couplings).
Since we ignore  the dynamical details 
of the underlying theory, we do not make specific assumptions
about the values of these coefficients. We only impose (to simplify the analysis)
the natural hierarchy
\beq
a_i = \cO(1)~, \qquad \epsilon_i  \ll 1~,
\eeq
which holds in most scenarios. The small coefficients, $\epsilon_i$, are induced by radiative corrections and are naturally of $\cO(0.1)$.
In this limit the series in Eq.~(\ref{eq:y_eff}) 
are dominated by the terms not suppressed by the $\epsilon_i$. 
As we will discuss below, relaxing the assumption about the smallness
of the $\epsilon_i$ complicates the reconstruction of the GUT Yukawa 
couplings in terms of the measured fermion masses, but it does
not modify in a significant way their hierarchical structure. 

The two matrices $\lambda_5$ and $\lambda_5^{\prime}$ have 
eigenvalues that are definitely not proportional,\footnote{~Neglecting the higher-order terms,
these matrices are diagonalized by bi-unitary transformations to go 
to the quark and lepton mass eigenstate basis.} 
but they have a similar hierarchical structure. 
A strong hierarchy is also exhibited by the eigenvalues of $\yten$.
For this reason the higher-order combinations 
of $\ysf$ and $\yten$ appearing in Eq.~(\ref{eq:y_eff}) 
have a very limited impact on the first two generations
(compared to the linear terms): 
in absence of $\ysig$ the GUT relation 
$m_e/m_\mu = m_d/m_s$ is almost unchanged even  for $\epsilon_i = \cO(1)$. 
In principle, one could hope to modify the $m_e/m_\mu = m_d/m_s$ relation
with a suitable choice of the $\ynu^T \ynu^* \yfive$ terms 
($\ynu$ contains large mixing angles and is not very hierarchical).
However, these terms have a non-trivial impact in Eqs.~(\ref{eq:y_eff})
only if the normalization of $\ynu$ is sufficiently high. As we will
show in section~\ref{sect:FCNC}, in this case one would induce too large FCNCs 
in the down-quark sector. As a result, the most natural solution
to break the  $m_e/m_\mu = m_d/m_s$ relation is the introduction 
of the additional $\bar5 \times 10$ spurion $\ysig$. 
Motivated by the phenomenological success of the first 
GUT relation in Eq.~(\ref{eq:m_ratios}) \cite{BEGN}, 
throughout this paper we will assume 
that the maximal eigenvalue of $\lambda_5$ is much larger than 
the maximal eigenvalue of $\lambda_5^{\prime}$.



The effective Majorana mass matrix of low-energy neutrinos, 
obtained by integrating out the $N_i$ fields and other sources 
of $U(1)_{\rm LN}$ breaking, has the following general structure:
\beq
m_\nu = \frac{v^2}{M_{\nu}} \left[ \ynu^T \frac{ M_{\nu}}{M_R} \ynuph
+ \epsilon_{1\nu}^{(\prime\prime)}  \ysf \yten^\dagger (\ysf)^T +\ldots \right]~.
\label{eq:m_nu}
\eeq
Making contact with our previous papers \cite{MLFV,CG,MLFV_lepto},
we have indicated with $M_{\nu}$ the average right-handed 
neutrino mass, which we assume to be the dominant (lighter) 
source of total lepton-number breaking. The $\epsilon_i$
in Eq.~(\ref{eq:m_nu}) take into account the possible effects of 
additional $U(1)_{\rm LN}$ breaking terms (and additional breaking of the 
discrete symmetry $L_L\to i L_L$, $h\to -h$), normalized 
to $1/M_\nu$. Also in this case we assume $\epsilon_{i} \ll 1$,
such that the standard see-saw mechanism dominates $m_\nu$. 

To display explicitly the strength of the flavor changing neutral
transitions, it is convenient to transform to the quark and lepton 
fields mass-eigenstate basis. To
this end we make unitary transformations on the spinor fields as
follows:
\begin{gather}
u_L\to V_{u_L} u_L~,u_R\to V_{u_R} u_R~,
d_L\to V_{d_L} d_L~,d_R\to V_{d_R} d_R~, \\
e_L\to V_{e_L} e_L~,e_R\to V_{e_R} e_R~,
\nu_L\to V_{\nu_L} \nu_L
\end{gather}
These transformations are chosen to diagonalize the mass matrices,
\begin{gather}
\ybu = V_{u_R}^\dagger \yuuph V_{u_L}^{\phantom{\dagger}}\,,~ 
\ybd = V_{d_R}^\dagger \yddph V_{d_L}^{\phantom{\dagger}}\,,~ 
\ybe = V_{e_R}^\dagger \yeeph V_{e_L}^{\phantom{\dagger}}\,,~
{\bar m_{\nu}} = V_{\nu_L}^T m_{\nu}^{\phantom{\dagger}} V_{\nu_L}^{\phantom{T}}\,,~  
\end{gather}
where $\ybu,\ybd,\ybe$ and $\bar m_{\nu}$ are diagonal. Note that
$V_{u_R}^*=V_{u_L}^{\phantom{\dagger}}$, which follows from 
$\yuu^T=\yuuph$, a feature in GUT models which does not 
generally hold in non-unified theories.   Flavor changing
interactions occur when the operators in the effective Hamiltonian 
fail to remain flavor diagonal after these unitary transformations
are performed. The well known CKM matrix appears in the transformation
of a charged current, $\bar u_L \gamma^\mu d_L\to \bar u_L V_{u_L}^\dagger V_{d_L}^{\phantom{\dagger}}
\gamma^\mu d_L$ so $\vckm= V_{u_L}^\dagger V_{d_L}^{\phantom{\dagger}}$. Similarly, $\upmns=V_{e_L}^\dagger
V_{\nu_L}^{\phantom{\dagger}}$.

Note that without loss of generality we can choose a $SU(5)$ invariant basis
(or appropriate flavor rotations of the ${\bf \bar 5}$ and ${\bf 10}$
fermion fields) that have $\ydd$ diagonal. 
In this basis\footnote{~Here $I$ denotes the $3\times 3$ identity matrix.},
\begin{gather}
V_{d_L}=V_{d_R}=I \qquad  V_{u_L} = V_{\rm CKM}^\dagger  \qquad    V_{\nu_L} = V_{e_L} \, U_{\rm PMNS}
\end{gather}
so that the Yukawa matrices read 
\begin{align}
\yuu &= V_{\rm CKM}^T \, \ybu \, V_{\rm CKM}^{\phantom{T}} \\
\yee &=   V_{e_R}^{\phantom{\dagger}}  \, \ybe  \, V_{e_L}^\dagger  \label{eq:new_mix_ma}  \\
\ynn &=  \frac{M^{1/2}_R}{v}~R~ \left({\bar m_\nu}\right)^{1/2}~\udagpmns~V_{e_L}^\dagger~,
\end{align}
where $R$ is a complex orthogonal matrix~\cite{CasasI}. 
If we impose $M_R=M_{\nu}I$ then the last equation takes the form
$\ynn =  (M^{1/2}_\nu /{v})~H~ \left( {\bar m_\nu}\right)^{1/2}~\udagpmns~V_{e_L}^\dagger~$, 
where $H$ is orthogonal and hermitian. If we further impose that CP is conserved in the neutrino 
Yukawas, then we can set $H\to I$. In this limit, the mixing matrices 
necessary to describe all possible flavor-changing processes are 
the well-known $\vckm$ and $\upmns$, and the two additional matrices 
$V_{e_L}$ and $V_{e_R}$ which
control the diagonalization of $\yee$ in 
the basis where $\ydd$ is diagonal.

\section{FCNC transitions}
\label{sect:FCNC}

\subsection{Old Mixing Structures}
\label{sect:oldmix}
The low energy effects of new physics that results from integrating
out fields at a scale $\Lambda$ is described by an effective
Hamiltonian constructed from SM fields. It contains an infinite series
of operators of ever increasing dimension, starting from dimension
five, with inverse powers of $\Lambda$ included to give correct
engineering dimensions. Hence the low energy effects are suppressed by
powers of the low energy scale over $\Lambda$, and the higher the
dimension of the operators the higher the power of the suppression. We
will concentrate on operators of dimension six at most.  The basic
building blocks of operators that may produce flavor changing neutral
effects are fermion bilinears. 
In the non-GUT framework analyzed in Ref.~\cite{MFV,MLFV}, 
all the relevant FCNC amplitudes are constructed in terms 
of the following bilinear combinations of fermion 
fields:
\bea
 &{\rm quarks}:& \quad
 {\bar Q}_L \yuu^\dagger \yuuph Q_L\,,  \qquad  {\bar d}_R \yddph \yuu^\dagger \yuuph Q_L\,, 
\label{eq:oldQFCNC} \\
&{\rm leptons:}& \quad
{\bar L_L} \ynn^\dagger \ynnph  L_L\,, \qquad  {\bar e_R} \yeeph \ynn^\dagger \ynnph L_L~.
\label{eq:oldLFCNC} 
\eea
In these bilinears the Dirac indices are implicit and free, that is,
they are not contracted. 
We have only retained combinations up to cubic order in the Yukawa
couplings. Some combinations, like $\bar Q_L \ydd^\dagger\yddph Q_L$, have not
been listed because they produce smaller effects than the one listed,
but they could become relevant in a two Higgs doublet model at large
$\tan \beta$. 

Consider the first operator in Eq.~(\ref{eq:oldQFCNC}). It contains
the down-type quark term which when transformed to the quark mass
eigenstate basis becomes,
\beq
\bar d_L \yuu^\dagger \yuuph d_L \to \bar d_L V_{d_L}^{\dagger}\yuu^\dagger \yuuph V_{d_L}d_L={\bar d}_L \Delta^{(q)}d_L 
\eeq
where $\Delta^{(q)}$ is completely specified in terms of the up-type
quark masses and the CKM mixing angles:
\beq \XQFV_{ij} \equiv  V_{\rm CKM}^\dagger \,  \ybu^2  \, V_{\rm CKM} =
\frac{m^2_t}{v^2} (\vckm)^*_{3i} (\vckm)_{3j} +\cO(m_{c,u}^2/m_t^2)~.
\label{eq:DFCNC_q}
\eeq
Similarly the first operator in Eq.~(\ref{eq:oldLFCNC}) contains the
charged lepton transition term 
\beq
\bar e_L \ynn^{\dagger}\ynnph e_L \to \bar e_L V_{e_L}^{\dagger}\ynn^{\dagger}\ynnph V_{e_L} e_L=\bar e_L \Delta^{(l)}e_L ~,
\eeq
with 
\beq
\XLFV \equiv  U_{\rm PMNS} \, 
(\bar{m}_{\nu})^{1/2}  
\, R^\dagger  
\, \frac{M_R}{v^2} \, R \, (\bar{m}_{\nu})^{1/2} \, U_{\rm PMNS}^\dagger ~ .
\eeq
Here the matrix $\Delta^{(l)}$ can be expressed in terms of neutrino
masses and mixings if one makes the simplifying assumption of
degenerate heavy neutrinos and approximate $CP$
invariance~\cite{MLFV}.  In this case
\begin{multline}
\XLFV_{ij} =
\frac{M_\nu}{v^2} \, \Big[ m_{\nu_1} \, \delta_{ij}  
+  (\upmns)_{i2} (\upmns^*)_{j2} \, (m_{\nu_2} - m_{\nu_1})    \\
+  (\upmns)_{i3} (\upmns)^*_{j3} \, (m_{\nu_3} - m_{\nu_1}) \Big]=
\frac{M_\nu \sqrt{\Delta m^{2}_{\rm atm}} }{v^2}  \, \xlfv_{i j}~, \quad 
\end{multline}
where, for later convenience we have introduced the `reduced' couplings 
$\xlfv_{i j}$ which are free of the (unknown) overall normalization.
More explicitly, using the PDG notation of the PMNS matrix (with 
the convention $s_{13}\geq0$), 
assuming maximal mixing for the atmospheric neutrinos and denoting with
$s$ and $c$ sine and cosine of the solar mixing angle, we find
\begin{align}
\xlfv_{2 1} & =  
 \frac{1}{\sqrt{2 \Delta m^{2}_{\rm atm}}  } \, \left[
s \, c  \, 
(m_{\nu_2} - m_{\nu_1})  
\pm s_{13} \, 
(m_{\nu_3} - m_{\nu_1})   
\right]~,
\label{eq:XLFV_12}
\\
\xlfv_{3 1} & =  
 \frac{1}{\sqrt{2 \Delta m^{2}_{\rm atm}}  } \, \left[
- s \, c  \, 
(m_{\nu_2} - m_{\nu_1})  
\pm s_{13} \, 
(m_{\nu_3} - m_{\nu_1})   
\right]~,
\label{eq:XLFV_13}
\\
\xlfv_{3 2} & =  
 \frac{1}{2 \sqrt{\Delta m^{2}_{\rm atm}}  } \, \left[
-  c^2  \, 
(m_{\nu_2} - m_{\nu_1})  
+   
(m_{\nu_3} - m_{\nu_1})   \right]~,
\label{eq:XLFV_23}
\end{align}
where the $+$ and $-$ signs correspond to $\delta=0$ and $\pi$,
respectively. In the  normal hierarchy case ($\nu_1$ is the lightest neutrino), one has:
\beq
m_{\nu_2} - m_{\nu_1}  ~ \stackrel{m_{\nu_1} \to 0}{\longrightarrow} ~
\sqrt{\Delta m^{2}_{\rm sol}}~,  \qquad\quad  
m_{\nu_3} - m_{\nu_1}  ~ \stackrel{m_{\nu_1} \to 0}{\longrightarrow} ~
\sqrt{\Delta m^{2}_{\rm atm}}~,
\eeq
while in the inverted hierarchy case ($\nu_3$ is the lightest neutrino)
\beq
m_{\nu_2} - m_{\nu_1} ~ \stackrel{m_{\nu_3} \to 0}{\longrightarrow} ~
\frac{\Delta m^2_{\rm sol}}{2 \sqrt{\Delta m^{2}_{\rm atm}}}~,   \qquad\quad  
m_{\nu_3} - m_{\nu_1} ~  \ \stackrel{m_{\nu_3} \to 0}{\longrightarrow} ~
 -  \sqrt{\Delta m^{2}_{\rm atm}}~.
\eeq

The key point which emerges by comparing Eqs.~(\ref{eq:XLFV_12})--(\ref{eq:XLFV_23})
and Eq.~(\ref{eq:DFCNC_q})  is the fact that  $\xlfv_{ij}$ 
is substantially less hierarchical than $\XQFV_{ij}$. 
In the 2--3 case, whose result is insensitive to the value of 
$s_{13}$ and is also very stable with respect to possible 
$CP$-violating parameters in $H$, we have 
\beq
\left| \xlfv_{3 2} \right| \approx  \frac{1}{2} \qquad {\rm vs.} \qquad
\left| \XQFV_{3 2} \right| \approx  0.04~.
\label{eq::XLFV_32_b}
\eeq
The difference is even more pronounced in the 1--2 case, where
\beq 
\left| \xlfv_{1 2} \right| \approx  {\rm max}\left[\frac{s_{13}}{\sqrt{2}},~
\frac{s c \sqrt{\Delta m^{2}_{\rm sol}} }{\sqrt{2 \Delta m^{2}_{\rm atm}}  } \right]
\approx 0.1
\qquad {\rm vs.} \qquad
\left| \XQFV_{1 2} \right| \approx  3\times 10^{-4}~. 
\label{eq::XLFV_12_b}
\eeq
In principle, there are fine-tuned scenarios where the two components
in Eq.~(\ref{eq:XLFV_12}) tend to cancel each other yielding smaller values of 
$|\xlfv_{1 2}|$. However, if we allow non-vanishing $CP$-violating  
parameters in the neutrino Yukawa coupling, then 
$\xlfv_{1 2}$ receives additional contributions 
proportional to $\sqrt{\Delta m^{2}_{\rm atm}}$ not suppressed by $s_{13}$ 
\cite{MLFV_lepto}: these new terms naturally increase the size of 
$|\xlfv_{1 2}|$, making these fine-tuned scenarios even more unlikely. 
Thus the estimate in (\ref{eq::XLFV_12_b}) can be considered as a 
fairly general lower bound on $|\xlfv_{1 2}|$.

\subsection{New Mixing Structures}
As anticipated, the restricted flavor group of the GUT framework 
allows more independent spurion combinations. 
In particular, potentially 
interesting effects arise from:
\bea
&{\rm quarks}:& \qquad  {\bar Q}_L (\yeeph \yee^\dagger)^T Q_L\,, \label{eq:newQLLFCNC} \\
&&   \qquad  {\bar d}_R \yee^T  (\yeeph \yee^\dagger)^T   Q_L\,,
 \qquad\!  {\bar d}_R  (\yeeph \ynn^\dagger \ynnph)^T   Q_L\,, \label{eq:newQRLFCNC} \\
&&    \qquad {\bar d}_R (\yee^\dagger \yeeph)^T d_R\,, 
 \qquad\quad {\bar d}_R (\ynn^\dagger \ynnph)^T d_R\,,  
        \label{eq:newQRRFCNC}   \\
&{\rm leptons:}& 
\qquad  \bar L_L(\yddph \ydd^\dagger )^T L_L\,,
\label{eq:newLLLFCNC} \\
&& \qquad  \bar e_R  (\yddph   \ydd^\dagger\yddph )^TL_L,\,\qquad \bar e_R\yuuph\yuu^\dagger\ydd^T L_L\,, 
\label{eq:newLRLFCNC} \\ 
&&\qquad {\bar e_R} \yuuph\yuu^\dagger e_R\,,\qquad\qquad\ \bar e_R (\ydd^\dagger \yddph )^Te_R\,, 
\label{eq:newLRRFCNC} 
\eea 
and terms obtained by the exchange $\ydd^{\phantom{T}} \leftrightarrow \yee^T$ in
any of the bilinears in (\ref{eq:oldQFCNC})--(\ref{eq:oldLFCNC}) and
(\ref{eq:newQLLFCNC})--(\ref{eq:newLRRFCNC}).  Note the unusual
structure, with right-handed down-type and charged-lepton fields not
necessarily suppressed by the corresponding Yukawa couplings.  To
display explicitly the strength of the flavor changing neutral
transitions we go to the quark and lepton mass eigenstate basis.
While the structures in~\eqref{eq:oldQFCNC} and \eqref{eq:oldLFCNC}
can be expressed solely in terms of the diagonal Yukawas and
the matrices $\Delta^{(q)}$ and $\Delta^{(l)}$, that is not the
case with the bilinears allowed by the restricted flavor group of the
GUT framework in \eqref{eq:newQLLFCNC}--\eqref{eq:newLRRFCNC}.

Consider, for example, the bilinear \eqref{eq:newQLLFCNC}. In the
mass-eigenstate basis it gives the flavor neutral bilinears
\beq
\bar u_L \big( V_{u_L}^\dagger V_{e_R}^*\ybe^2 V_{e_R}^T V_{u_L}\big)
u_L\,,
\quad\text{and}\quad
\bar d_L \big(  V_{d_L}^\dagger V_{e_R}^*\ybe^2 V_{e_R}^T V_{d_L} \big)d_L\,.
\label{eq:CC_ee}
\eeq
There is a new mixing matrix, $C\equiv V_{e_R}^T V_{d_L}$, in terms of
which these bilinears take the form
\beq
\bar u_L \big(\vckm C^\dagger \ybe^2 C \vdackm \big)  u_L\,,\quad\text{and}\quad
\bar d_L \big(C^\dagger \ybe^2 C \big)d_L\,.
\label{eq:CC_ee2}
\eeq

The complete list of new, independent mixing matrices  is
\begin{align}
C &= V_{e_R}^T V_{d_L}^{\phantom{T}}\\
G &= V_{e_L}^T V_{d_R}^{\phantom{T}}
\end{align}
As shown in section~\ref{sect:QLMasses}, these two matrices diagonalize $\yee$ 
in the basis where $\ydd$ is diagonal. Indeed $C\to I$ and $G\to I$
in the limit $\lambda_e^{T} \to \lambda_d$.
The flavor changing neutral bilinears that follow from
\eqref{eq:newQLLFCNC}--\eqref{eq:newLRRFCNC} can be readily expressed
in terms of $C$ and $G$. For example, the spurions in $\bar d_L\otimes d_L$
of \eqref{eq:newQLLFCNC}, including those that follow by replacing
$\yee\to\ydd^T$ are
\begin{equation}
\label{eq:dLdLstructures}
 C^\dagger \ybe^2 C\,,\quad C^\dagger \ybe G \ybd\,,\quad\text{and}\quad
\ybd G^\dagger \ybe C\,.
\end{equation}
Note that no new independent mixing matrices arise involving the
transformations on $u$-quarks, because $V_{u_R}^*$ is not independent
of $V_{u_L}$ and the latter can be traded for $V_{d_L}\vdackm$. Had we
kept the non-symmetric contribution 
to $\yuu$ proportional to $\lambda_{10}^{'}$ (see Eq.~\eqref{tenprime}) one
more structure, the matrix $V_{u_L}^TV_{u_R}^{\phantom{T}}$, would arise. 

The two new mixing structures arise from different alignment between
the mass matrices of charge $-1/3$ quarks and charged leptons. As seen
in Eq.~\eqref{eq:y_eff},  the misalignment, due to $\lambda_5'$ and
the higher order terms, is small compared to the largest
eigenvalue.   
This allows one to determine the texture of the matrices $C$ and $G$
by proceeding in two steps: (i) first diagonalize the upper $2\times
2$ block of $\ydd$ and $\yee^T$ via bi-unitary block-diagonal
transformations; (ii) then diagonalize the remaining structure using
perturbation theory, through unitary matrices of the form
$U_{\epsilon} = I + \hat{\epsilon}$, with $\hat{\epsilon}^\dagger = -
\hat{\epsilon}$ and $\hat{\epsilon} \sim O(m_\mu/m_\tau)$. 
It follows that the matrices $C$ and $G$ have a hierarchical structure, 
\begin{equation}
\label{eq:C-struct-given}
C=\begin{array}({c|c})
U_C & U_C \, \epsilon_C\\
\hline
- \epsilon_C^\dagger  & 1
\end{array} \ + \ O(\epsilon^2) \, ,
\end{equation}
where $U_C$ is a unitary $2\times2$ matrix and $\epsilon_C$ is a
$2\times1$ matrix with small components, of order $m_\mu/m_\tau $. The
structure of the matrix $G$ is analogous, obtained by replacing $U_G$
for $U_C$ and $\epsilon_G$ for $\epsilon_C$. To understand the
consequences of this structure consider, for example, the first
combination in \eqref{eq:dLdLstructures}. A $b\to s$ or $b\to d$
transition may involve the largest eigenvalue $\bar \lambda_\tau^2 $
but is accompanied by one power of $\epsilon_C\sim 5 \times 10^{-2}$. Had we
been ignorant about the structure of $C$ we would have been forced to
conclude that any one off diagonal component of $C^\dagger \ybe^2 C$
could be in principle as large as $\bar \lambda_\tau^2$.

\subsection{Phenomenological constraints from FCNC processes}
A substantial simplification on the possible new structures 
arises in the limit of a single light Higgs boson, or two Higgs bosons 
with similar vevs, where $(\bar \lambda_{e,d})_{33}/(\ybu)_{33} \sim m_{b,\tau}/m_t \ll 1$. 
This strong suppression factor, combined with the hierarchy 
of $C$ and $G$,  allows us to neglect all terms with at least two powers 
of $\lambda_{e,d}$. Note that this is not a good approximation in two Higgs
doublet models at large $\tan\beta$. 
As a result, the complete list of phenomenologically 
relevant bilinear structures to be added to `old' terms in (\ref{eq:oldQFCNC}) 
reduces to 
\beq
\ba{l}
{\bar d}_R (\ynn^\dagger \ynnph)^T d_R   \\
{\bar d}_R  (\yeeph \ynn^\dagger \ynnph)^T  Q_L \\
{\bar d}_R  (\ynn^\dagger \ynnph)^T \yddph Q_L \\
{\bar e_R} \yuuph\yuu^\dagger e_R \\
{\bar e_R} \yuuph\yuu^\dagger\ydd^T L_L \\
{\bar e_R} \yuuph\yuu^\dagger\yeeph L_L
\ea
\longrightarrow
\ba{l}
{\bar d}_R G^\dagger \left(\Delta^{(l)}\right)^T G d_R  \\
{\bar d}_R G^\dagger \left(\Delta^{(l)}\right)^T \ybe C d_L  \\
{\bar d}_R G^\dagger \left(\Delta^{(l)}\right)^T G \ybd d_L  \\
{\bar e_R} \left[C \Delta^{(q)} C^\dagger\right]^* e_R \\
{\bar e_R} \left[C \Delta^{(q)} \ybd G^\dagger\right]^* e_L \\
{\bar e_R} \left[C \Delta^{(q)} C^\dagger\right]^* \ybe e_L 
\ea
\label{eq:newLRRFCNC_r} 
\eeq
where the arrow denotes the expressions relevant to 
down-type quarks and charged leptons (in the corresponding 
mass-eigenstate basis). In the following 
we will analyze some phenomenological consequences of 
these new structures derived from $\Delta F=2$ quark processes 
and radiative FCNC decays.

First of all, it is clear that if the overall normalization 
of $\lambda_1$ is sufficiently small, we have no practical 
deviations from the non-GUT MFV scenario in the quark sector.
As remarked above, this conclusion does not hold with two-Higgs 
doublets and large $\tan\beta$: given that $|C_{13}|$
is parametrically larger than $|(\vckm)_{13}|$, 
if $(\ybe)_{33} \approx m_{\tau} \tan\beta/v$ 
is sufficiently large the bilinear in (\ref{eq:CC_ee2}), which has two
factors of $\yee$, 
can generate sizable non-standard FCNC contributions 
in $b\to d$ and $s\to d$ transitions.

In order to quantify how small the normalization 
of $\lambda_1$ should be not to affect quark FCNC transitions, we  
analyze $\Delta F=2$ processes, on which several 
precise experimental data are available [$\epsilon_K$ for 
$2\leftrightarrow 1$ mixing, $\Delta M_{B_d}$ and $\cA_{CP}(B_d \to\psi K^0)$ 
for $3\leftrightarrow 1$, $\Delta M_{B_s}$ for $3\leftrightarrow 2$]. 
Here the relevant dimension-six effective Hamiltonian describing 
new-physics effects (renormalized around the 
electroweak scale) contains 
\beq
\cH^{\Delta F=2} = \frac{1}{\Lambda^2} 
\left[  c_{1} ({\bar Q}_L \yuu^\dagger \yuuph Q_L)^2 +  c_{2}
({\bar d}_R (\ynn^\dagger \ynnph)^T d_R)^2 \right]~,
\label{eq:HDF2}
\eeq
where $\Lambda$ denotes the effective scale of new physics. 
The main virtue of the MFV hypothesis is to allow new physics 
close to the electroweak scale, as expected by a natural 
stabilization of the electroweak symmetry-breaking sector. 
Given the normalization of  $\cH^{\Delta F=2}$, the natural value 
of $\Lambda$ for $c_{i}=\cO(1)$ is $\Lambda \lsim 10$~TeV, if the new  physics 
that generates these operators is related to the 
solution of the hierarchy problem.\footnote{~The effective 
FCNC operators generated within the SM by integrating out the top-quark and the heavy 
gauge-boson fields, correspond to an effective scale 
$\Lambda_0 = \sin \theta_W  M_W/\alpha_{\rm em} \approx  2.4 ~{\rm TeV}~$ \cite{MFV}.}
In the case of the first operator in $\cH^{\Delta F=2}$, 
which survives also in the non-GUT case,  
present data implies the 95~\%CL bound $\Lambda > 5.7$~TeV 
(for $c_{1}=1$)~\cite{Bona}, which is consistent with this 
naturalness assumption.  
We estimate that the second term in $\cH^{\Delta F=2}$ is consistent with
present experimental limits provided it does not exceed the largest
allowed value of the first one (reached for $\Lambda = 5.7$~TeV with
$c_{1}=1$).  Therefore, with $c_{2}=1$, the new mixing term must
satisfy the condition
\beq
\left| \left(G^\dagger \left(\XLFV\right)^T G\right)_{i \not= j} \right| 
=  \frac{M_\nu \sqrt{\Delta m^{2}_{\rm atm}} }{v^2} 
\left| \left(G^\dagger \left(\xlfv\right)^T G\right)_{i \not= j} \right| 
\quad \lsim  \quad 
\left| \XQFV_{i \not= j} \right|~. 
\label{eq:l1b}
\eeq 
which constraints the overall normalization of $\ynn$.  
The unknown structure of $G$ combined with the unknown hierarchy 
of the neutrino mass spectrum does not allow a precise evaluation 
of the l.h.s.~of (\ref{eq:l1b}). However, the variability range is quite 
limited thanks to the non-hierarchical structure of $\XLFV$
(see section~\ref{sect:oldmix}). Barring accidental 
cancellations, we find
\beq
0.1 \lsim \left| \left(G^\dagger \left(\xlfv\right)^T G\right)_{i \not= j} \right| \lsim 1~.
\label{eq:DLmix}
\eeq
The most conservative scenario (as far 
as the extraction of lower bounds on $M_\nu$ is concerned) is obtained 
for the normal hierarchy of the neutrino spectrum and $m_{\nu_1}\to 0$, 
where we obtain\footnote{~In the case of normal hierarchy and  $m_{\nu_1}\to 0$,
the 1--2 mixing term in (\ref{eq:DLmix}) is $\cO(\left(\xlfv\right)_{12})$.
For inverted hierarchy and large 1--2 entries in $G$, the mixing can become
larger resulting in more stringent constrains on $M_\nu$.}
\beq
M_\nu  \lsim  
\left\{ \ba{c} 6 \times 10^{13}~{\rm GeV}\quad {\rm from~2-3~mixing,}
  \\ 2\times10^{12}~{\rm GeV} \quad {\rm from~1-2~mixing.}  
\ea \right. 
\label{eq:Mnu_upper}
\eeq 

Interestingly, the most stringent upper limit on $M_\nu$ in (\ref{eq:Mnu_upper})
is very close but not incompatible with the lower limit $M_\nu \gsim 10^{12}$~{\rm GeV}
derived in \cite{MLFV_lepto} as the condition for successful leptogenesis
(with  MFV and degenerate right-handed neutrinos). Since the 
most stringent constraint in (\ref{eq:Mnu_upper}) arises from  
1--2 mixing, if $M_\nu \sim 10^{12}~{\rm GeV}$ 
we could expect deviations from the non-GUT MFV relations 
of Ref.~\cite{MFV} in rare $K$ decays, while we should not 
see appreciable effects in $B$ physics. This is in contrast 
with the expectation of non-MFV GUT models such as the one 
analyzed in \cite{MM}, where the large 2--3 mixing angle 
in the neutrino sector is used to advocate large FCNC 
contributions in $b\to s$ transitions.

Once we impose that the normalization of $\ynn$ is 
low enough to pass the constraints from $\Delta F=2$ processes,
the two LR quark bilinears in (\ref{eq:newLRRFCNC_r}) 
do not play a significant role in $b\to s\gamma$.
The situation is much more interesting for the radiative 
lepton decays $l_i \to l_j \gamma$,
whose relevant effective Hamiltonian (written below the scale 
of $SU(2)_L\times U(1)_Y$ breaking) contains 
\beq
\Delta {\cal H}^{\rm \Delta F=1}_{\rm eff}
=\frac{v}{\Lambda^2}  \bar e_R \left[~c'_{1}~\yeeph \ynn^\dagger\ynnph 
~+~c'_{2}~\yuuph\yuu^\dagger \lambda_e
~+~c'_{3}~\yuuph\yuu^\dagger\lambda_d^T~\right] \sigma^{\mu \nu} e_L F_{\mu \nu } ~.
\eeq
Here the new mixing terms induced by the quark Yukawa couplings 
could dominate over the non-GUT FCNC structures 
analyzed in \cite{MLFV}. 

We start recalling the upper bounds on $M_\nu$ following 
from the non-observation of the  $l_i \to l_j \gamma$
processes in the non-GUT case ($c'_2=c'_3=0$). In this 
case the parametric  pattern of the leading FCNC couplings is 
\beq
\left| \left(\ybe \, \Delta^{(l)} \right)_{ij} \right| \sim 
\frac{M_\nu \sqrt{\Delta m^{2}_{\rm atm}} }{v^2}
\times \left\{ 
 \ba{l}  
   \cO(1) \bar{\lambda}_\tau     \\
   \cO(0.1) \bar{\lambda}_\tau     \\
   \cO(0.1) \bar{\lambda}_\mu  
 \ea
\qquad 
 \ba{l}
    (\tau_R \to \mu_L) \\
    (\tau_R \to e_L) \\
    (\mu_R  \to e_L) 
 \ea 
\right.
\label{eq:LFV_ll}
\eeq
The dominant constraint comes from the experimental 
limit $\mathcal{B}(\mu\to e\gamma) < 1.2\times 10^{-11}$. 
Imposing the same conditions adopted in the quark 
sector ($\Lambda \lsim 10$~TeV with $c'_{1}=1$), 
we find $M_\nu \lsim 2 \times 10^{12}$~GeV~\cite{MLFV}.
Surprisingly enough, this limit is very similar to the one 
imposed by $\Delta F=2$ quark mixing. This implies that 
$\mu\to e\gamma$ should be close to  its present exclusion 
limit: if $M_\nu \sim 10^{12}$~GeV all the three terms 
in $\Delta {\cal H}^{\rm \Delta F=1}_{\rm eff}$ can yield 
comparable contributions to the $\mu\to e\gamma$  rate, while 
if $M_\nu < 10^{12}$~GeV the two quark-induced terms 
start to dominate. The only way to suppress  $\mathcal{B}(\mu\to e\gamma)$
below $10^{-12}$ is to push  the new-physics scale 
to  high values above $10$ TeV.

The scenario where we can neglect the first term in 
$\Delta {\cal H}^{\rm \Delta F=1}_{\rm eff}$ is quite similar, 
but not identical, to the one considered in Ref.~\cite{BH} in the 
context of supersymmetry.
We fully recover the flavor-mixing pattern of Ref.~\cite{BH} 
in the limit where we neglect $\ysig$ and $C,G \to I$.
In the more general case, taking into account the 
hierarchical structure of $C$, $G$ and $\Delta^{(q)}$ 
the leading mixing terms in $\Delta {\cal H}^{\rm \Delta F=1}_{\rm eff}$
have the following parametric structure:
\beq
\left|\left(C \Delta^{(q)} \ybd G^\dagger\right)_{ij}\right| \sim 
\left|\left(C \Delta^{(q)} C^\dagger \ybe\right)_{ij} \right| \sim 
\frac{m_t^2}{v^2}  \times \left\{ 
 \ba{l}  
   \epsilon \bar{\lambda}_\tau     \\
   \epsilon \bar{\lambda}_\tau     \\
   \epsilon^2 \bar{\lambda}_\mu  
 \ea
\stackrel{C,G \to I}{\longrightarrow} 
 \ba{l}  
   \lambda^2           \bar{\lambda}_\tau     \\
   \lambda^3    \bar{\lambda}_\tau     \\
   \lambda^5  \bar{\lambda}_\mu  
 \ea
\quad 
 \ba{l}
    (\tau_L  \to \mu_R) \\
    (\tau_L \to e_R) \\
    (\mu_L \to  e_R) 
 \ea 
\right.
\label{eq:LFV_qq}
\eeq
where $\lambda\sim (\vckm)_{12} \sim 0.2$ and $\epsilon\sim \lambda^2$ denotes the parametric size
of both $(\vckm)_{23}$ and $C_{i3}$ and $G_{i3}$ for $i=1,2$. The more
pronounced hierarchy of the couplings relevant to $\tau$ FCNC decays in
(\ref{eq:LFV_qq}) vs.~(\ref{eq:LFV_ll}) implies that the present limit
on $\mathcal{B}(\mu\to e\gamma)$ does not forbid values of $\mathcal{B}(\tau\to
\mu,e\gamma)$ above $10^{-9}$ (as concluded in \cite{MLFV} looking only at
the non-GUT terms).

\section{Conclusions}
 
In this work we have implemented the principle of the Minimal Flavor Violation   
in Grand Unified theories,  focusing our attention on the  $SU(5)$ gauge group. 

Since quarks and leptons of a given family (including $\nu_R$) are 
grouped in three  irreducible representations  of SU(5) ($\bar {\bf 5}$, ${\bf 10}$ and 
${\bf 1}$),   the flavor symmetry group  $G_F = SU(3)_{\bar 5} \times SU(3)_{10} \times SU(3)_{1}$
of  the gauge Lagrangian  is smaller compared to the standard case.
We identified the irreducible sources of 
 $SU(3)_{\bar 5} \times SU(3)_{10} \times SU(3)_{1}$-breaking  with the minimal 
 set of couplings  that leads to a consistent 
fermion mass spectrum at low energy. This can be done both with minimal 
and non-minimal Higgs content, e.g.~introducing new Higgs fields transforming 
as a ${\bf 45}$.  

One noteworthy  consequence of the smaller flavor group is that the effective FCNC 
couplings are not determined anymore in terms of the diagonal fermion mass matrices 
and the CKM and PMNS mixing matrices. 
Due to the different alignment of the down-quarks and charged leptons mass matrices,  
two new mixing structures appear, of which only the hierarchical texture is known. 
The presence  of new mixing structures precludes  the possibility of deriving  precise 
relations among the rates for different family transitions. Only an order-of-magnitude pattern 
can be identified. Despite this, a number of reasonably firm 
phenomenological consequences  can be deduced: 
\begin{itemize}
\item There is a well defined limit in which the standard MFV scenario 
for the quark  sector is  fully recovered: small $\tan \beta$ and $M_\nu \ll 10^{12}$ GeV.
For $M_\nu \sim  10^{12}$ GeV and small $\tan \beta$, deviations from the standard MFV pattern 
can be expected in rare Kaon decays but  not in $B$ physics. Ignoring fine-tuned 
scenarios, $M_\nu \gg  10^{12}$~GeV is excluded by the present constraints 
on quark FCNC transitions. Independently from the value of $M_\nu$, 
deviations from the standard MFV pattern can appear both in $K$ and in $B$ physics
for $\tan\beta \gsim m_t/m_b$. 

\item 
Contrary to the non-GUT MFV framework,  
the rate for $\mu \to e \gamma$ (and other leptonic FCNC) cannot be 
arbitrarily suppressed by lowering the  average mass $M_\nu$ of the heavy  $\nu_R$. 
For $M_\nu \lsim 10^{12}$ GeV, the GUT-induced  
contribution (controlled mainly 
by the top-quark Yukawa coupling and the CKM matrix) 
sets in. This implies that for 
values of the new physics scale $\Lambda \lsim 10$ TeV 
the  $\mu \to e \gamma$  rate is within reach of the experimental searches at MEG~\cite{MEG} . 

\item Improved experimental information on $\tau \to \mu \gamma$ and $\tau \to e \gamma$ would 
be a powerful  tool in discriminating  the relative size of the 
standard  MFV contributions  (see Eqs. \ref{eq:LFV_ll}))
versus  the characteristic GUT-MFV contributions (see Eqs.~(\ref{eq:LFV_qq})), 
due to the different hierarchy 
pattern among $\tau \to \mu$, $\tau \to e$, and $\mu \to e$ transitions. 
  
\end{itemize}

\section*{Acknowledgment}
We thank Enrico Nardi for useful discussions.  VC, BG and MBW
acknowledge the theory group at INFN-LNF for hospitality during the
Spring Institute 2006.  This work was supported by the U.S.~Deartment
of Energy under grants DE-FG03-97ER40546 and DE-FG03-92ER40701.  The
work of VC was carried out under the auspices of the National Nuclear
Security Administration of the U.S. Department of Energy at Los Alamos
National Laboratory under Contract No. DE-AC52-06NA25396.

\end{document}